\documentclass[twoside]{ilcws07}

\usepackage[latin1]{inputenc}

\usepackage[dvips]{graphicx,epsfig,color}

\usepackage{wrapfig,rotating}

\usepackage{amssymb,amsmath,array}

\usepackage{epsfig}

\begin{document}


\title{Loops for ILC}



\author{Matthias Steinhauser
\vspace{.3cm}\\
Institut f{\"u}r Theoretische Teilchenphysik,
Universit{\"a}t Karlsruhe (TH)\\
76128 Karlsruhe, Germany
}



\maketitle

\begin{abstract}
This contribution summarizes the on-going activities connected to 
the evaluation of higher order radiative corrections in the context of
a future international linear collider (ILC).
\end{abstract}


\section{Introduction}

The purpose of this contribution is two-fold. The primary task is to
present a summary of the activities discussed 
in the parallel session ``loops'' of the Linear Collider
Workshop (LCWS) 2007 at DESY in Hamburg.
As a second aim we try to provide an overview of
higher order corrections performed in the context of the ILC. 
It is clear that a
brief review like the present one can not be complete and has to be 
restricted to the most important issues.
For further related activities we want to refer to the summaries of
the Top/QCD, Higgs, SUSY and extra dimensions parallel sessions which
can also be found in these 
proceedings~\cite{Hew-Kan}.


\section{Bhabha scattering}

Let us in a first step discuss the activities in the context of the
next-to-next-to-leading order (NNLO) corrections to the Bhabha
scattering which serves as an important luminosity monitor for 
basically all electron-positron colliders. The uncertainty in the
luminosity enters into many 
observables and thus needs to be determined with the highest possible
precision. This is in particular true for the Giga-$Z$ option of the
ILC. 

In the recent years various groups have started the NNLO calculation to the
Bhabha scattering which constitutes a highly non-trivial task since
next to the kinematic variables $s$ and $t$ also the mass of the
electron, $m_e$, has to be kept non-zero. As far as the
dependence of the scattering cross section on $m_e$ is concerned, it
is only necessary to keep the logarithmic dependence and neglect the 
terms suppressed by $m_e^2/s$.

The calculation of the cross section $\sigma(e^+e^-\to e^+e^-)$ 
for $m_e=0$ has been
performed in Ref.~\cite{Bern:2000ie}.
In Ref.~\cite{Penin:2005kf} this result has been used in order to perform a
matching to the case where the infra-red singularities are regularized
by a photon mass and the collinear ones by the electron mass. In this
way the NNLO corrections for the purely photonic correction to the
Bhabha scattering could be obtained. A similar approach has been elaborated
in Ref.~\cite{Becher:2007cu} where, however, the infrared divergences are still
regularized dimensionally leading to more flexibility,
in particular in view of applications within QCD (see also
Ref.~\cite{Mitov:2006xs}). 

The fermionic corrections which are defined by the presence of a
closed lepton loop have been considered in Ref.~\cite{Bonciani:2004gi} for the
case of an electron loop. Recently, the results for a muon and tau
have been obtained in Ref.~\cite{Actis:2007gi}. In the approach used in this
paper a reduction of the full multi-scale problem to master integrals is
performed. Afterwards the latter are expanded in the desired kinematical
limit. The results of Ref.~\cite{Actis:2007gi} have been confirmed in
Ref.~\cite{Becher:2007cu} (see also Ref.~\cite{Bonciani:2007eh}).

There are various further contributions which are still missing to complete
the NNLO corrections. Among them is the
computation of the one-loop corrections where an additional photon is
radiated.
Progress on the evaluation of the underlying five-point
integrals have been presented at this workshop~\cite{Riemann}.


\section{NLO corrections to multi-particle production}

In the recent years there have been important developments concerning
the techniques for one-loop calculations involving many external legs
(see, e.g., Ref.~\cite{Ellis:2006ss} and references therein). However, 
many of the proposed methods still have to prove their applicability to real
processes.

Up to date there are only two groups who performed a full one-loop
calculation to a realistic $2\to4$ process. In Ref.~\cite{Denner:2005es} the
process $e^+e^-\to4f$ has been considered and in Ref.~\cite{Boudjema:2005rk}
electroweak corrections to $e^+e^-\to \nu\bar{\nu} HH$ have been
obtained using the {\tt GRACE} system (see, e.g., Ref.~\cite{Belanger:2003sd}).

In a contribution~\cite{Schwinn} to the present workshop an
effective-theory approach has been introduced, based on a
double-expansion in the fine structure constant and the ratio of
width and mass of the $W$ boson. In the threshold region, which is the
validity range of the effective theory, good agreement with the
results of Ref.~\cite{Denner:2005es} has been
found for the cross section of the process $e^+e^-\to
\mu^-\bar{\nu}_\mu u\bar{d}$.

In contribution~\cite{Yasui} new developments for the 
{\tt GRACE} system has been discussed. Among them 
there is an interface to {\tt FORM}, the
implementation of one-loop calculations in the MSSM and the proper
treatment of infrared divergences in QCD processes. Furthermore, there
is a new attempt to obtain octuple (or even a higher) precision in the
numerical routines.


\section{Sudakov logarithms}

With the ILC it will be possible to consider the corrections of virtual $W$ and
$Z$ bosons to exclusive reactions like the production of two quarks or two $W$
bosons. Since the center-of-mass energy is significantly higher than the
masses of the gauge bosons a conceptually new phenomenon occurs: in each
loop-order quadratic logarithms of the form $\ln^2(s/M_{W/Z}^2)$ arise
which can easily lead to corrections of order 30\% at one and 5\% at two
loops. For recent papers dealing with this topic we refer to
Refs.~\cite{Jan-Den-Kuh}.

At LCWS07 a recent calculation has been presented~\cite{Jantzen} which deals with
the complete two-loop NLL corrections to processes like $f_1 f_2\to f_3\ldots f_n$
involving $n$ fermions. Furthermore, a new approach has been discussed
which allows for the introduction of finite quark masses for the final state
particles. 


\section{NNLO calculation to $e^+e^-\to 3$~jets}

An accurate determination of the strong coupling can be obtained 
by the measurement of 
the 3-jet cross section in $e^+e^-$ annihilation. Currently the error on
$\alpha_s$ from this method is dominated by the theoretical uncertainties
which is mainly due to the unknown NNLO corrections to $e^+e^-\to 3$~jets.

There are basically three ingredients contributing to $e^+e^-\to
3$~jets: (i) the two-loop virtual corrections, (ii) the one-loop
corrections to the real radiation 
of a parton, and (iii) the double real radiation which involves five partons in
the final state. The individual contributions are known since many
years (see contribution~\cite{Gehrmann} to this workshop). However, up
to very recently  
a proper combination of the individual pieces has not been achieved.
The main reason for this are the infrared divergences inherent to the
contributions (i), (ii) and (iii) which only cancel in the proper
combination. In the recent 
years different approaches have been developed which are either based on the
construction of appropriate subtraction terms or on direct numerical
integration. The latter essentially relies on sector decomposition.

In Ref.~\cite{GehrmannDe Ridder:2007bj} the first
physical NNLO result has been presented 
for the thrust distribution defined through $T=\mbox{max}_{\vec{n}}
\frac{\sum_{i=1}^n |\vec{p}_i\cdot\vec{n}|}
{\sum_{i=1}^n |\vec{p}_i|}$. 
The corrections turn out to be moderate leading to a significant reduction of
the theoretical uncertainty on the thrust distribution.


\section{Four-loop integrals}

At the forefront of multi-loop calculations one also has to mention the
contributions to four-loop vacuum integrals and four-loop massless two-point
functions. The former integrals, often also denoted as ``bubbles'', are
reduced with the help of the so-called
Laporta-algorithm~\cite{Laporta} to master 
integrals. The latter are evaluated with various methods based, e.g., on
difference equations or on asymptotic expansion (see, e.g.,
Refs.~\cite{Schroder:2005va,Chetyrkin:2006dh}).

Two applications have been presented at the 
LCWS07. In the first one the four-loop
corrections to the $\rho$ parameter have been
studied~\cite{Sturm,Schroder:2005db,Chetyrkin:2006bj,Boughezal:2006xk}.
The new terms induce a shift in the $W$ boson mass of about 2~MeV which is of
the same order as the anticipated accuracy reached with the GIGA-$Z$ option of
the ILC. The latter is estimated to $6$~MeV.

The second application~\cite{Steinhauser} concerns the extraction of precise
values for the charm and bottom quark masses which in the $\overline{\rm MS}$
are given by~\cite{Kuhn:2007vp} $m_c(m_c)=1.286(13)$~GeV and
$m_b(m_b)=4.164(25)$~GeV. The analysis performed in Ref.~\cite{Kuhn:2007vp} is
based on 
improved experimental data to $\sigma(e^+e^-\to\mbox{hadrons})$ and new
four-loop contributions to the photon polarization
function~\cite{Chetyrkin:2006xg,Boughezal:2006px}.

Also the four-loop massless two-point functions have various applications where
the most important one is the order $\alpha_s^4$ correction to the
cross section $\sigma(e^+e^-\to\mbox{hadrons})$ (see, e.g.,
Ref.~\cite{Baikov:2006nb}
for a recent publication). Their evaluation is based on Baikov's
method~\cite{Baikov:2003zq} 
where the reduction to master integrals is established via an integral
representation for the coefficients of the individual master integrals. The
parameter integrals are solved in the limit of large space-time dimension, $d$. 
Due to the fact that the coefficients are rational functions of $d$ it is
possible to reconstruct the exact $d$ dependence, provided sufficient
expansion terms are available.


\section{Further loops}

There have been four further contributions 
which shall be mentioned in this Section.

New two-loop electroweak corrections to the partial decay width of the Higgs
boson to bottom quarks have been presented in
contribution~\cite{Butenschoen} (see also Ref.~\cite{Butenschoen:2007hz}). 
Although the new terms are enhanced by a factor $(G_F m_t^2)^2$ the change of
the partial decay rate is tiny and amounts to only $0.05$\%.

In contribution~\cite{Seidel} new three-loop corrections to the relation
between the $\overline{\rm MS}$ and on-shell quark mass have been presented.
In contrast to the previously known terms an additional mass scale from closed
quark loop is allowed~\cite{Bekavac:2007tk} where the main phenomenological
applications are charm quark corrections to the bottom quark mass. The
reduction of all occurring integrals leads to 27 master integrals which involve
two mass scales. They have been computed both with the help of the
Mellin-Barnes and the differential equation technique.

The production of a Higgs boson at LHC 
in the so-called vector-boson fusion channel is
very promising for its discovery. At LO in perturbation theory the gauge
bosons are radiated off the quarks and combine in order to produce the
Higgs boson. There is no colour exchange between the quarks and thus it is
expected that two jets are observed at high rapidity whereas the decay
products of the Higgs boson can be found at low rapidity. Thus, it is
possible to apply cuts which allow for a huge suppression of the background.
The exchange of colour between the quark lines occurs for the first time at
NNLO.
In contribution~\cite{Weber} the NNLO corrections originating from 
squared one-loop amplitudes with gluons in the initial state have been
considered. Preliminary results have been presented which show that the
numerical effect is small if the so-called ``vector-boson fusion'' cuts are
applied.

In contribution~\cite{Hoang} (see also Ref.~\cite{Fleming:2007qr}) a new
method has been proposed to extract a precise top quark mass value from jet
observables. It is based on a sequence of effective field theories which
allows to derive a factorization theorem for the top quark invariant mass
spectrum. The factorization theorem allows for a separation of perturbative
and non-perturbative effects which in turn is the basis of the extraction of
the so-called ``jet mass''. 

For the evaluation of higher order quantum corrections it is crucial
to have appropriate tools which facilitate the
calculations~\cite{Harlander:1998dq}. As far as one-loop corrections
are concerned one should 
mention {\tt FeynArts}~\cite{Hahn:2000kx} and {\tt
  FormCalc}~\cite{Hahn:1998yk} which have been applied to a variety of 
processes in the electroweak sector of the Standard Model but also in
its extensions. Beyond one-loop the programs in general aim for 
specific tasks of the whole calculation. E.g., 
the program {\tt AIR}~\cite{Anastasiou:2004vj}
implements the Laporta algorithm, the {\tt Mathematica} codes {\tt
  AMBRE}~\cite{Gluza:2007rt} and {\tt MB}~\cite{Czakon:2005rk}
can be used to evaluate Feynman integrals with the 
Mellin-Barnes method,
and the program {\tt exp}~\cite{Harlander:1997zb} allows for the
application 
of an Euclidian asymptotic expansion for a given hierarchy in the
mass scales involved in the problem.
A tool which nowadays is indispensable in higher order calculations is
the algebra program {\tt FORM}~\cite{Vermaseren:2000nd} enabling large
computations 
in a quite effective way. Also its parallel versions, {\tt
  ParFORM}~\cite{Tentyukov:2006pr} and {\tt
  TFORM}~\cite{Tentyukov:2007mu}, have proven
to substantially extend the capability of {\tt FORM}.




\vspace*{.3em}
\noindent
{\bf Acknowledgments}
\\
This work was supported by the DFG through SFB/TR~9.
We thank the Galileo Galilei Institute for Theoretical Physics for the
hospitality and the INFN for partial support during the completion of
this write-up.
\vspace*{-.3em}


\begin{footnotesize}

\end{footnotesize}


\end{document}